\documentclass[twocolumn,showpacs,preprintnumbers,amsmath,amssymb]{revtex4}
\usepackage{amssymb}
\usepackage{amsmath}
\usepackage{txfonts}

\usepackage{graphicx}
\usepackage{dcolumn}
\usepackage{bm}

\begin{document}

\title{QED-based Optical Bloch Equations without electric dipole approximation: A model for a two-level atom interacting with a monochromatic X-ray laser beam}

\author{Wen-Zhuo Zhang and Wu-Ming Liu}
\affiliation{Beijing National Laboratory for Condensed Matter Physics, Institute of Physics, Chinese Academy of Sciences, Beijing 100190, P.R. China}

\begin{abstract}
We derive a set of optical Bloch equations directly from the minimal-coupling Hamiltonian density of the bound-state quantum electrodynamics (bound-state QED). Such optical Bloch equations are beyond the former widely-used ones due to that there is no electric dipole approximation on the minimal-coupling Hamiltonian density of the bound-state QED. Then our optical Bloch equations can describe the time evolution of a two-level atom in a monochromatic light of arbitrary wavelength, which are suitable to study the spectroscopy and the Rabi oscillations of two-level atoms in X-ray laser beams since that the wavelength of X-ray is close to the atom to make the electric dipole approximation invalid.
\end{abstract}


\maketitle

\section{Introduction}

Optical Bloch equations (OBEs) are the most widely used time-dependent equations for light-atom interaction in quantum optics \cite{QO1,QO2}, with which the stimulated absorption, stimulated emission, and spontaneous emission of light by atoms are well described, and the spectroscopy and Rabi problem of atoms can be studied well. The OBEs play the central role in both the semi-classical models and the quantum models such as Jaynes-Cummings model \cite{JC}. In these models, the interaction Hamiltonian is $H_I=-e\mathbf{D\cdot E}$, which is restricted by the \emph{electric dipole approximation} (EDA). Such approximation is a combination of two individual approximations. The first one is the dipole approximation, which requires that the radius of an atom $\mathbf{r}$ ($\sim0.1 nm$) is much less than the wavelength of the light $\mathbf{\lambda}$,
\begin{equation}
2\pi\cdot\mathbf{\frac{r}{\lambda}}=\mathbf{k\cdot r}\ll1.
\end{equation}
The second approximation neglects the coupling between the atoms and the magnetic field of the light $\mathbf{B}$ due to that it is much smaller than the coupling between the atoms and the electric field $\mathbf{E}$ of the light. Therefore, these widely-used OBEs in quantum optics textbooks are only valid for the interaction between atoms and long-wavelength light.

Recent realization of the free-electron lasers that work on X-ray wavelengths \cite{FEL,X-ray} (although only femtosecond pulses are available at present) provides an opportunity for the experimental study the time evolution of atoms in monochromatic X-ray beam. With the steady-state solutions ($t\gg\tau$, with $\tau$ being the lift-time of the excited state) and instantaneous solutions ($t<\tau$) of OBEs one can study the spectroscopy of atoms in continues-wave (CW) laser and the Rabi problem of atoms in ultrafast laser pulse (such as femtosecond pulses), respectively. Such OBEs are different from the X-ray scattering theories \cite{XRS} due to that the OBEs are time-dependent which study the evolution of the system, while the X-ray scattering theories study the time-independent initial and final states of the system. However, the wavelength of X-ray ($\sim0.01 nm$ to $10 nm$) does not satisfy the dipole approximation condition ($\mathbf{r\ll\lambda}$), which makes the widely-used OBEs in quantum optics textbooks invalid in studying the interaction between atoms and X-ray laser. Then a new set of OBEs beyond the electric dipole approximation is required for the spectroscopy and Rabi oscillations of atoms in X-ray laser beams.

Quantum electrodynamics (QED), which is the relativistic quantum field theory for the electromagnetic interaction, studies the light-matter interactions at the fundamental level \cite{QED}. Bound-state QED is the extension of QED which studies the interaction between photons and bound electrons in atoms with the language of QED. The early bound-state QED models are developed by Furry \cite{Furry}, Salpeter and Bethe \cite{BS1,BS2}, as well as Gell-Mann and Low \cite{GL} to study the bound-state problem in quantum field theory. In the past years, many bound-state QED models were developed to correct the electrons' energy levels in atoms and ions \cite{Buch,correction1,correction2}, including few-electron atoms \cite{FE1,FE2} and many-electron atoms \cite{ME1,ME2,ME3,ME4,ME5,ME6}. Since the interaction Hamiltonian in the bound-state QED models does not have the EDA, it can describe the interaction between the bound electrons in atoms and the light at all wavelength. Then it is interesting to ask a question that whether a new set of OBEs can be derived directly form the interaction Hamiltonian in bound-state QED? If so, such OBEs can get rid of the EDA in order to describe the interaction between atoms and light of arbitrary wavelength, especially to meet the recent development of X-ray laser.

In this paper, we start from the minimal-coupling Hamiltonian density between the bound-state Dirac field and the free electromagnetic field in bound-state QED to derive the new style of OBEs. Our new OBEs can describe a two-level atom interacting with a monochromatic light of arbitrary wavelength, which meet the requirement of studying the spectroscopy and the Rabi oscillations of two-level atoms in X-ray laser beams. Our paper is arranged as follows: In Section II, we present the interaction Hamiltonian for a light-atom interacting system form the interaction Hamiltonian density between bound-state Dirac field and free electromagnetic field in bound-state QED, and derive the time-dependent equations of the ground and excited states of a two-level atom in interaction picture. In Section III, we involve the density matrix of the two-level atom and the spontaneous emission of its excited state into the time-dependent equations in last section, and derive the OBEs for a two-level atom interacting with monochromatic light. We present how the new OBEs can describe the spectroscopy and the Rabi oscillations of a two-level atom in X-ray laser. Section V is the conclusion of this paper, which also gives the discussions on possible extensions of such new OBEs.

\section{Time-dependent equations for a two-level atom in a monochromatic light form bound-state QED}

In the interaction picture, the interaction Hamiltonian density of quantum electrodynamics (QED), which is the minimal-coupling between free Dirac field and free electromagnetic field, is
\begin{equation}
\mathcal{H}_I=-\mathcal{L}_I=-e\tilde{\varphi}\gamma^{\mu} A_{\mu}\varphi,
\end{equation}\label{I-QED}
with $\varphi$ being the Dirac field, $\gamma^{\mu}$ being the Dirac matrices, $A_{\mu}$ being the covariant four-potential of the electromagnetic field, and $e$ being the elementary charge. The minimal-coupling Hamiltonian density for the interaction between the bound-state Dirac field and free electromagnetic field in bound-state QED has the similar form \cite{Furry,Buch}
\begin{equation}
\mathcal{H}_I^{B}=-e\tilde{\varphi}_B A^{\mu}\varphi_B.\label{I-B}
\end{equation}
Here $\varphi_B$ is the bound-state Dirac field which describes the bound electrons in atoms, and $A^{\mu}$ is the four-potential of the free electromagnetic field.

Quantization of the bound-state Dirac field, $\varphi_B$, is different to it of the free Dirac field, $\varphi$. The reason is that the eigenstates of the bound electrons are the discrete wave functions of energy and angular momentum rather than the continues wave-functions of kinetic energy and momentum for free electrons. Since no positrons exist in atoms, we simplify the Furry picture \cite{Furry} by removing the creation and annihilation operators of positrons, and only preserving the creation and annihilation operators of electrons. Then the bound-state Dirac field for the valence electron in a two-level atom can be canonically quantized as
\begin{equation}
\begin{aligned}
\varphi_B&=\sum_p\sum_{m=e,g}\beta_{p,m}\phi_m(x)e^{\frac{i}{\hbar}p\cdot x}e^{\frac{i}{\hbar}E_mt}b_m,\\
\tilde{\varphi}_B&=\sum_p\sum_{m=e,g}\beta^*_{p,m}\phi^*_m(x)e^{-\frac{i}{\hbar}p\cdot x}e^{-\frac{i}{\hbar}E_mt}b^\dag_m.\label{QB}
\end{aligned}
\end{equation}
Here, $\phi_m(x)$ is the space-dependent wave-function of the bound electron in the state $|m>$ ( with $|e>=|1_e,0_g>$ being the excited state and $|g>=|0_e,1_g>$ being the ground state ), $p$ is the external linear momentum of the bound-state Dirac field, $E_n=\hbar\omega_m$ is the total energy of the bound-state Dirac field in the state $m$, and $\beta_{p,m}$ is the normalization parameter. $b^\dag_m$/$b_m$ is the creation/annihilation operator of the bound electrons in the state $|m>$, which obeys the anti-commutation relation $\{b^\dag_m,b_{m'}\}=\delta_{mm'}$, and $\{b^\dag_m,b^\dag_{m'}\}=\{b_m,b_{m'}\}=0$.

The canonical quantized monochromatic electromagnetic field in Coulomb gauge can be written
\begin{equation}
A^{\mu}=\sum_{\lambda=1}^2\varepsilon\hat{\epsilon}_{k,\lambda}(a^\dag e^{-ik\cdot x-\omega_lt}+ae^{ik\cdot x+\omega_lt}),\label{QA}
\end{equation}
with $k$ being its wave-vector, $\omega_l$ being its frequency, $\varepsilon=\left(\frac{\hbar c^2}{2V\omega_k}\right)^{1/2}$ being the normalization coefficient, $\hat{\epsilon}_{k,\lambda}$ being the vector polarization, and $a^\dag/a$ being the creation/annihilation operator of photons which obeys the commutation relation $[a,a^\dag]=1$, and $[a^\dag,a^\dag]=[a,a]=0$. Since the angular momentum of a bound electron is fixed at a certain energy level in atoms, we assume that the polarization of the electromagnetic field (circular or linear polarization) just fits the requirement of the angular momentum difference ($\sigma^{\pm}$ or $\pi$ transitions) between $|e>$ and $|g>$. Then $\sum_{\lambda=1}^2\hat{\epsilon}_{k,\lambda}$ can be replaced by $\hat{\epsilon}_k$ for such two-level system.

After the quantization of both bound-state Dirac field and monochromatic electromagnetic field, we can integrate Eq.~(\ref{I-B}) in the three-dimensional space to obtain the time-dependent interaction Hamiltonian $H_I(t)$.
\begin{equation}
\begin{aligned}
H_I(t)&=\int dx^3 \tilde{\varphi}_B A_{\mu}\varphi_B\\
&=-e\varepsilon\beta^*_e\beta_g ab^\dag_eb_g e^{-i(\omega_e-\omega_g-\omega_l)t}\\
&\times\int\phi^*_e(x)\hat{\epsilon}_k\phi_g(x)e^{\frac{i}{\hbar}(p_e-p_g)\cdot x+ik\cdot x}dx^3\\
&-e\varepsilon\beta^*_g\beta_e a^\dag b^\dag_gb_e e^{i(\omega_e-\omega_g-\omega_l)t}\\
&\times\int\phi^*_g(x)\hat{\epsilon}_k\phi_e(x)e^{\frac{i}{\hbar}(p_g-p_e)\cdot x-ik\cdot x}dx^3.\label{Ht}
\end{aligned}
\end{equation}
Here we adopt the rotating-wave approximation to ignore the counter-rotating frequency parts $\exp{[\pm i(\omega_e-\omega_g+\omega_l)t]}$, which is an widely used rational approximation in quantum optics \cite{QO1,QO2}. The integrations of $\int\phi^*_e(x)\hat{\epsilon}_k\phi_g(x)dx^3$ and $\int\phi^*_g(x)\hat{\epsilon}_k\phi_e(x)dx^3$, which show the transition probability between $|e>$ and $|g>$, can be calculated with the wave-function, $\phi_e(x)$ and $\phi_g(x)$. Here we define
\begin{equation}
\begin{aligned}
\Delta&=\omega_l-(\omega_e-\omega_g),\\
q&=-e\varepsilon\beta^*_e\beta_g\int\phi^*_e(x)\hat{\epsilon}_k\phi_g(x)e^{\frac{i}{\hbar}(p_e-p_g)\cdot x+ik\cdot x}dx^3,\\
q^*&=-e\varepsilon\beta^*_g\beta_e\int\phi^*_g(x)\hat{\epsilon}_k\phi_e(x)e^{\frac{i}{\hbar}(p_g-p_e)\cdot x-ik\cdot x}dx^3,\label{int}
\end{aligned}
\end{equation}
The the spatial integrations ($q$ and $q*$) are the key differences between the conventional interaction Hamiltonian with EDA and the minimal coupling Hamiltonian in Eq.~(\ref{Ht}). In the interaction Hamiltonians which are based on EDA \cite{QO1,QO2}, the spatial integrations ($q$ and $q*$) are not considered due to that the wavefunctions of electrons are localized to the spatial point where the atom is ($x=0$), then such integration are replaced by localized electric-dipole transition matrix elements ($<e|\mathbf{r}|g>$ and $<g|\mathbf{r}|e>$) at $x=0$. In the minimal coupling Hamiltonian, Eq.~(\ref{Ht}), the integrations can be directly calculated with the wavefunctions of electrons, which is free from the wavelength of light. Then Eq.~(\ref{Ht}) can be used to derive the OBEs with arbitrary light wavelength. With Eq.~(\ref{int}) the time-dependent interaction Hamiltonian $H_I(t)$ can be rewritten as
\begin{equation}
H_I(t)=q^*e^{-i\Delta t}a^\dag b^\dag_gb_e+qe^{i\Delta t}ab^\dag_eb_g.\label{HI}
\end{equation}
This is the bound-state QED based interaction Hamiltonian for a two-level atom interacting with photons.

\begin{figure}
\includegraphics[width=50mm]{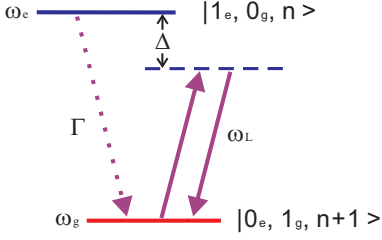}
\caption{(Color online) Transitions of a two-level atom in a quantized monochromatic light. Solid lines with arrows denote the stimulated excitation and emission of a monochromatic photon between the ground state $|0_e,1_g,n+1>$ and the excited state $|1_e,0_g,n>$, where $e$ and $g$ are the excited and ground state, respectively. $n$ is the particle number state of the photons, $\omega_l$ is the frequency of the photons, and $\Delta=\omega_l-(\omega_e-\omega_g)$ is the detuning of the photons to the energy difference between $|0_e,1_g>$ and $|1_e,0_g>$. Dotted line with an arrow denote the spontaneous emission of the excited state, with $\Gamma$ being the spontaneous emission rate}\label{level}
\end{figure}

Following the standard procedure in quantum optics textbooks \cite{QO1,QO2}, we can define the time-dependent particle number state $|\Psi(t)>$ of the system in the interaction picture,
\begin{equation}
|\Psi(t)>=c_e(t)|1_e,0_g,n>+c_g(t)|0_e,1_g,(n+1)>,\label{state}
\end{equation}
where $c_e(t)$ is the probability amplitude of the atom in the excited state $|1_e,0_g>$, $c_g(t)$ is the probability amplitude of the atom in the ground state $|0_e,1_g>$, and $n$ is the number of the photons. The Shr\"{o}dinger-like equations for $H_I(t)$ and $|\Psi(t)>$ in the interaction picture is
\begin{equation}
i\hbar\frac{\partial}{\partial t}|\Psi(t)>=H_I(t)|\Psi(t)>.\label{IP}
\end{equation}
Then with Eq.~(\ref{HI}), (\ref{state}), and (\ref{IP}), we can obtain the time-dependent equations of $c_e(t)$ and $c_g(t)$,
\begin{equation}
\begin{aligned}
\dot{c}_e(t)&=-\frac{i}{\hbar}q\sqrt{n+1}e^{i\Delta t}c_g(t),\\
\dot{c}_g(t)&=-\frac{i}{\hbar}q^*\sqrt{n+1}e^{-i\Delta t}c_e(t).\label{Rabi}
\end{aligned}
\end{equation}
This coupled set of equations describes the Rabi oscillation between the two level, $|1_e,0_g,n>$ and $|0_e,1_g,(n+1)>$, with the Rabi frequency
\begin{equation}
\begin{aligned}
\Omega=\frac{2q\sqrt{n+1}}{\hbar},\\
\Omega^*=\frac{2q^*\sqrt{n+1}}{\hbar}.
\end{aligned}
\end{equation}

The set of two equations in Eq.~(\ref{Rabi}) is an ideal model where the spontaneous emission of the excited state is not considered. For a real system of a two-level atom interacting with monochromatic light, the spontaneous emission of the excited state cannot be ignored. The density matrix of the system is necessary to combine Eq.~(\ref{Rabi}) and the spontaneous emission into a set of equations, which are just the optical Bloch equations (OBEs).

\section{Optical Bloch Equations for a two-level system beyond the electric dipole approximation}

To derive the OBEs, we need to introduce the density matrix for the two-level atom interacting with monochromatic light at first. The density matrix is
\begin{equation}
\rho=\begin{pmatrix}
\rho_{ee} & \rho_{eg} \\
\rho_{ge} & \rho_{gg}
\end{pmatrix}=
\begin{pmatrix}
c^*_e(t)c_e(t) & c^*_e(t)c_g(t) \\
c^*_g(t)c_e(t) & c^*_g(t)c_g(t)
\end{pmatrix},
\end{equation}
with $\rho_{ee}=c^*_e(t)c_e(t)$ being the probability of the atom in the excited state $|e>=|1_e,0_g>$, $\rho_{gg}=c^*_g(t)c_g(t)$ being the probability of the atom in the ground state $|g>=|0_e,1_g>$, and $\rho_{eg}=\rho^*_{ge}=c^*_e(t)c_g(t)$ being the off-diagonal elements which determine the transition between $|g>$ and $|e>$. An obvious relation is that $\rho_{ee}+\rho_{gg}=1$.

After introducing the density matrix, we need to include the spontaneous emission of the excited state $|e>=|1_e,0_g>$. The standard theory of spontaneous emission is the Weisskopf-Wigner theory \cite{WW}, which presents that the spontaneous emission is the result of the interaction between the two level atom at the excited state $|e>$ and the vacuum state of multi-mode electromagnetic field. The former interaction Hamiltonian in Weisskopf-Wigner theory is also based on the electric dipole approximation (EDA). In principle, we can rewrite the Weisskopf-Wigner theory with the minimal-coupling Hamiltonian of the bound-state QED
\begin{equation}
H_I(t)=\sum_k(q^*_ke^{-i\Delta t}a^\dag b^\dag_gb_e+q_ke^{i\Delta t}ab^\dag_eb_g),
\end{equation}
and the initial state of the system
\begin{equation}
|\Psi(t)>=c_e(t)|1_e,0_g,0>+\sum_kc^k_g(t)|0_e,1_g,1_k>.
\end{equation}
Such new form of Weisskopf-Wigner theory can get rid of the EDA. However, this work is worth to do in the future and we do not to present it here. In this paper, we adopt a constant $\Gamma$ as the spontaneous emission rate of the excited state $|e>=|1_e,0_g,n>$ (then the lifetime of the excited state becomes $1/\Gamma$), which is the widely-used approach in quantum optics \cite{QO1,QO2}.

Now with Eq.~(\ref{Rabi}) and $\Gamma$, we can write the time-evolution equations for every element of the density matrix, which are
\begin{equation}
\begin{aligned}
\dot{\rho}_{ee}(t)&=\dot{c}^*_e(t)c_e(t)+c^*_e(t)\dot{c}_e(t)-\Gamma\rho_{ee}\\
&=\frac{i}{\hbar}q^*\sqrt{n+1}e^{-i\Delta t}\rho_{ge}-\frac{i}{\hbar}q\sqrt{n+1}e^{i\Delta t}\rho_{eg}-\Gamma\rho_{ee},\\
\dot{\rho}_{gg}(t)&=\dot{c}^*_g(t)c_g(t)+c^*_g(t)\dot{c}_g(t)+\Gamma\rho_{ee}\\
&=-\frac{i}{\hbar}q\sqrt{n+1}e^{i\Delta t}\rho_{eg}+\frac{i}{\hbar}q^*\sqrt{n+1}e^{-i\Delta t}\rho_{ge}+\Gamma\rho_{ee},\\
\dot{\tilde{\rho}}_{eg}(t)&=(\dot{c}^*_e(t)c_g(t)+c^*_e(t)\dot{c}_g(t))e^{i\Delta t}+(i\Delta-\frac{\Gamma}{2})\tilde{\rho}_{eg}\\
&=\frac{i}{\hbar}q^*\sqrt{n+1}(\rho_{gg}-\rho_{ee})+(i\Delta-\frac{\Gamma}{2})\tilde{\rho}_{eg},\\
\dot{\tilde{\rho}}_{ge}(t)&=[\dot{\rho}_{eg}(t)]^*\\
&=\frac{i}{\hbar}q^*\sqrt{n+1}(\rho_{ee}-\rho_{gg})-(i\Delta+\frac{\Gamma}{2})\tilde{\rho}_{ge}.\label{OBE}
\end{aligned}
\end{equation}
Here $\tilde{\rho}_{eg}=\rho_{eg}e^{i\Delta t}$, and $\tilde{\rho}_{ge}=\rho_{ge}e^{-i\Delta t}$. The equations in Eq.~(\ref{OBE}) are our bound-state QED based OBEs for a two-level atom interacting with monochromatic light of arbitrary wavelength. We obtain these equations from the interaction Hamiltonian, Eq.~(\ref{HI}), which is directly derived from the interaction Hamiltonian density between a bound electron and a monochromatic electromagnetic field in bound-state QED. Therefore, Eq.~(\ref{OBE}) is the set of OBEs beyond the EDA. Comparing with the former OBEs that derived from Jaynes-Cummings model \cite{JC}, the brief change of our OBEs is that the Rabi frequency under the EDA is replaced by the $iq/\hbar$, where $q$ is the spatial integration of all space-dependent wave-functions of both photons and bound electrons (see Eq.~(\ref{int})). This is the reason why our bound-state QED based OBEs are free from any spatial approximation, including the EDA.

\begin{figure}
\includegraphics[width=90mm]{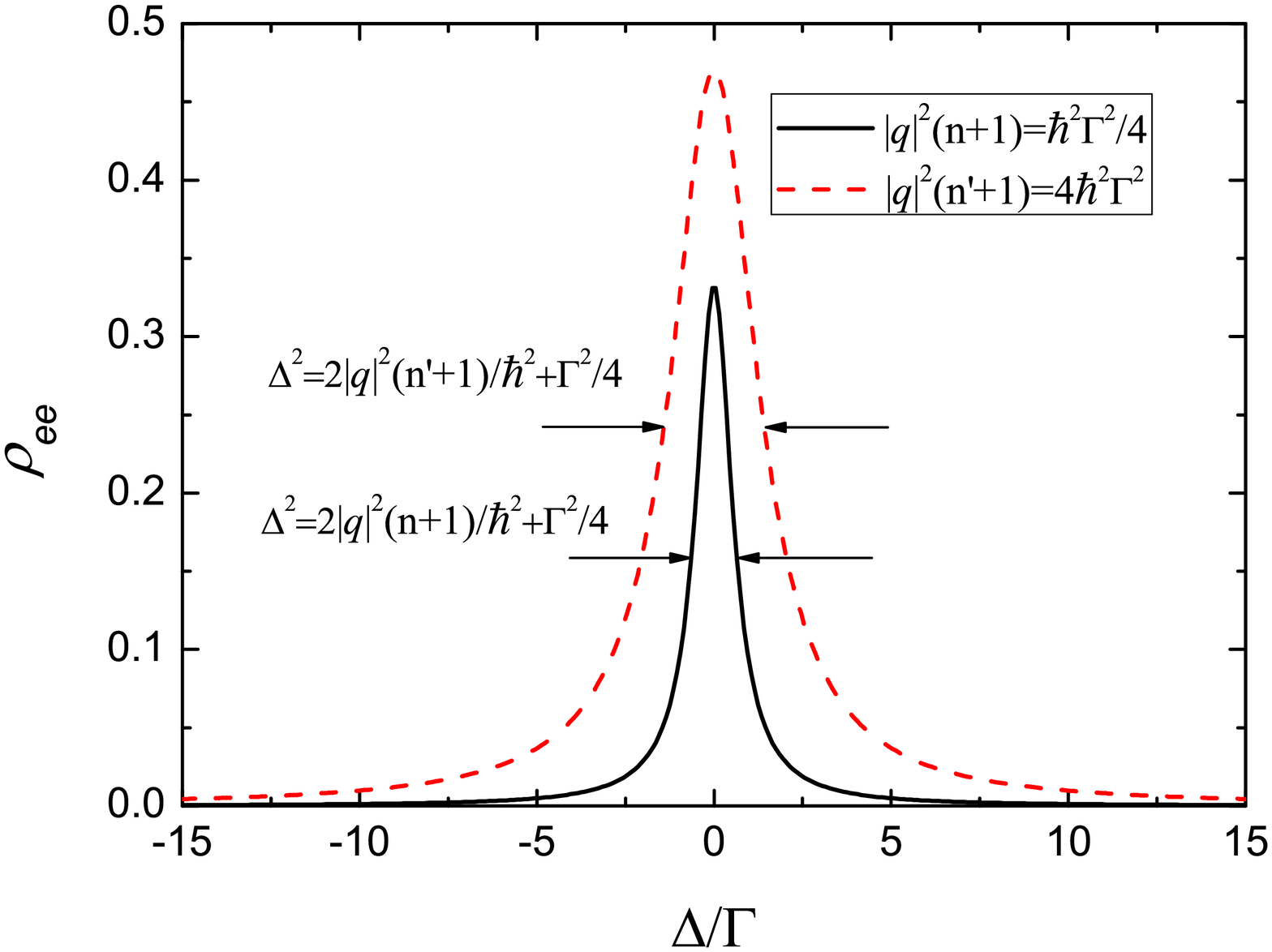}
\caption{(Color online) The variation of the excited state $\rho_{ee}$ of the two-level atom with $\Delta/\Gamma$ in the steady-state limit ($t\gg1/\Gamma$), with $\Delta$ being the detuning of the light and $\Gamma$ being the spontaneous emission rate. The solid (black) line is for photon number $n=\hbar^2\Gamma^2/4|q|^2-1$, and the dashed (red) line is for the photon number $n'=2\hbar^2\Gamma^2/|q|^2-1$. Since the intensity of the monochromatic light is proportional to its photon number $n$ in the particle number representation, a larger $n$ leads to a larger maximum value and a larger full-width-at-half-maximum (FWHM) of the $\rho_{ee}$}.\label{absorption}
\end{figure}

A very useful solution of Eq.(\ref{OBE}) is the steady-state solution of $\rho_{ee}$ under the condition $\dot{\rho}_{ee}(t)=\dot{\rho}_{gg}(t)=\dot{\tilde{\rho}}_{eg}(t)=\dot{\tilde{\rho}}_{ge}(t)=0$. The steady-state solutions of $\rho_{ee}$ and $\rho_{gg}$ are
\begin{equation}
\begin{aligned}
\rho_{ee}&=\frac{4|q|^2(n+1)/(\hbar^2\Gamma^2)}{1+8|q|^2(n+1)/(\hbar^2\Gamma^2)+4\Delta^2/\Gamma^2},\\
\rho_{gg}&=1-\rho_{ee}\\
&=\frac{1+4|q|^2(n+1)/(\hbar^2\Gamma^2)+4\Delta^2/\Gamma^2}{1+8|q|^2(n+1)/(\hbar^2\Gamma^2)+4\Delta^2/\Gamma^2}.
\end{aligned}
\end{equation}
Fig.~\ref{absorption} shows the varying of $\rho_{ee}$ with $\Delta/\Gamma$ under two different photon numbers $n$ and $n'$. The distribution of $\rho_{ee}$ is a Lorentz line-shape, which determines the amplitude and width of the absorption signal of the monochromatic light by a two-level atom in laser spectroscopy. The resonance point of the light and the two-level atom is at $\Delta=0$, where $\rho_{ee}$ has its maximum value $\frac{4|q|^2(n+1)}{\hbar^2\Gamma^2+8|q|^2(n+1)}$. We can see such maximum value of $\rho_{ee}$ is always less than $0.5$ due to that the $n\rightarrow\infty$ limit of $\rho_{ee}$ is just $0.5$ at $\Delta=0$ . This means that the population of the excited state $\rho_{ee}$ is always smaller than it of the ground state $\rho_{gg}$ in a steady-state two-level atom. Besides, $n$ and $|q|^2$ also determine the full-width-at-half-maximum (FWHM) of the absorption signal. In the condition $|q|^2(n+1)\ll\hbar^2\Gamma^2$, the FWHM of the absorption signal is approximately $\Gamma$, which means that the absorption signal of a weak-intensity light has the natural line-width $\Gamma$ of the excited state. When the light intensity (proportional to $n+1$) increases, the FWHM of the absorption signal is broaden to the value $\sqrt{4|q|^2(n+1)/\hbar^2+\Gamma^2}$, which is called power broadening in laser spectroscopy. Such steady-state solution of $\rho_{ee}$ is also proportional to the radiation force $F=\hbar\Gamma\rho_{ee}$ from the monochromatic light to the atom, which is very important in the laser cooling of atoms.

\begin{figure}
\includegraphics[width=90mm]{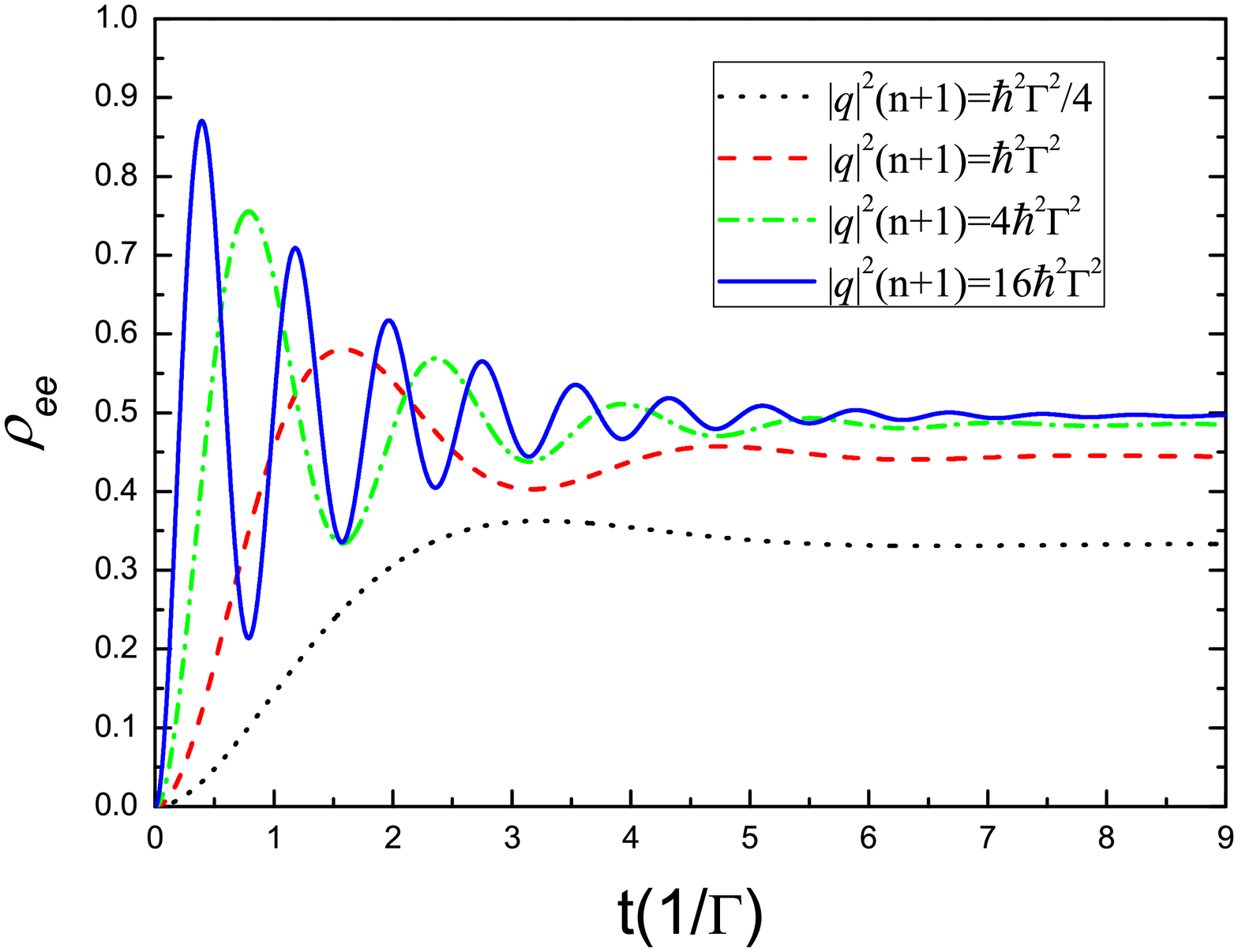}
\caption{(Color online) Evolution of the excited state population of $\rho_{ee}(t)$ between $t=0$ and $t=9/\Gamma$ under different values of $(n+1)|q|^2$ (with $n$ being the number photons which is proportional to the intensity of the light). $1/\Gamma$ is the lifetime of the excited state $|1_e,0_g,n>$. $\Delta=0$. The dotted (black) line, dashed (red) line, dash dotted (green) line, and solid (blue) line are the evolution of $\rho_{ee}(t)$ with $|q|^2(n+1)=\hbar^2\Gamma^2/4$, $|q|^2(n+1)=\hbar^2\Gamma^2$, $|q|^2(n+1)=4\hbar^2\Gamma^2$, and $|q|^2(n+1)=16\hbar^2\Gamma^2$, respectively. A larger $|q|^2(n+1)$ leads to a larger steady-state value of $\rho_{ee}(t\rightarrow\infty)$, larger Rabi oscillation frequency, and larger amplitude of $\rho_{ee}(t)$ before $t=9/\Gamma$.}\label{evolution}
\end{figure}

Another useful solution of Eq.~(\ref{OBE}) is the evolution of $\rho_{ee}(t)$ before the system reaches its steady-state. From Fig.~\ref{evolution} we can see such evolution of $\rho_{ee}(t)$ is also determined by the values of $n$ and $|q|^2$. When $|q|^2(n+1)=\hbar^2\Gamma^2/4$, the $\rho_{ee}(t)$ reaches its steady value $\rho_{ee}\approx3.3$ near $t=6\Gamma$, and has less than one oscillation period before it; when $|q|^2(n+1)=\hbar^2\Gamma^2$, the $\rho_{ee}(t)$ reaches its steady value $\rho_{ee}\approx4.4$ near $t=7\Gamma$, and has two Rabi oscillation periods before it; when $|q|^2(n+1)=4\hbar^2\Gamma^2$, the $\rho_{ee}(t)$ reaches its steady value $\rho_{ee}\approx4.8$ near $t=8\Gamma$, and has five obvious oscillation periods before it; when $|q|^2(n+1)=16\hbar^2\Gamma^2$, the $\rho_{ee}(t)$ reaches its steady value $\rho_{ee}\approx5.0$ near $t=9\Gamma$, and has ten obvious Rabi oscillation periods before it. The larger values of $|q|^2(n+1)$ cause (i) larger steady-state values of $\rho_{ee}(t\rightarrow\infty)$, (ii) larger damping time of $\rho_{ee}(t)$ before reaching steady-state, (iii) and larger Rabi oscillation frequencies as well as amplitudes of $\rho_{ee}(t)$ during the damping time. The value of $|q|^2$ is determined by the space-dependent wave-functions of the excited and ground states (see Eq.~(\ref{int})). When the Rabi oscillation disappears, the system reaches its final steady-state, where the phase parameter, $e^{\pm i\Delta t}$, disappears and the dephasing effect due to the spontaneous emission happens \cite{dephasing}. Therefore, the way to increase the Rabi oscillation frequencies and amplitudes, as well as the damping time of $\rho_{ee}(t)$ is to increase the photon number $n$, which is just to increase the intensity of the monochromatic light in practise.

Our new style of Optical Bloch Equations reach the similar conclusions to that of the Jaynes-Cummings model \cite{JC} in the case when the wavelength of the light is much larger than the radius of an atom (ultraviolet light, visible light, infrared light, and microwaves). In other cases when the wavelength of the light is close to or smaller than the radius of an atom (X-rays and Gamma rays), the Jaynes-Cummings model is not suitable due to that the EDA becomes invalid. However, our new OBEs can still reach the results of Fig.~\ref{absorption} and Fig.~\ref{evolution} without the EDA. Then our OBEs can be applied to describe the evolution of a two-level atom (with transition frequency being in the X-ray range) in a monochromatic in a X-ray laser beam. Such two-level atom could have the absorption spectroscopy of X-ray laser that described by Fig.~\ref{absorption}) in the steady-state ($t\gg1/\Gamma$), and have the Rabi oscillation that shown in Fig.~\ref{evolution} before the two-level system reaching its steady-state.

Another useful result in Fig.~\ref{evolution} is the population inversion ($\rho_{ee}>0.5$) of the two-level system during the first period of Rabi oscillations ($t\ll1/\Gamma$). Since the Rabi frequency and amplitude are proportional to $\sqrt{n+1}$, the time from $t=0$ to the maximum value of population inversion (we define it as $\Delta t$) can be shorten by increasing the intensity of the monochromatic light, while the $\rho_{ee}$ is also increased. Then the value of $\Delta t$ can be obtained from the evolution of $\rho_{ee}(t)$. This value is useful for the pulse pumped atomic X-ray laser system \cite{X-ray}. For example, a three-level system with two ground states $|g_1>, |g_2>$, and one excited state $|e>$ is prepared with all the papulation being at $|g_1>$. We can pump the population from the $|g_1>$ to $|e>$ to make a population inversion between $|e>$ and $|g_2>$, then the X-ray laser with the frequency between $|e>$ and $|g_2>$ can be made. In the case when the spontaneous emission rate from $|e>$ to $|g_1>$, $\Gamma_1$, is much larger than it from $|e>$ to $|g_2>$, $\Gamma_2$, the transition between $|e>$ to $|g_1>$ by a free-electron laser can be considered as a two-level system approximately, and can be described by the OBEs in Eq.~(\ref{OBE}). When the duration of an intense pulse from the free-electron laser is close to $\Delta t$, the population inversion between $|e>$ to $|g_1>$ happens (green line in Fig.~\ref{evolution}), which is the most effective population inversion between $|e>$ and $|g_2>$. Then from the OBEs in Eq.~(\ref{OBE}) one can calculate out $\Delta t$ to optimize the duration of the pump pulse in order to give an efficient pump rate for the atomic X-ray laser.

\section{discussion}

We consider a simple case that a two-level atom interacting with monochromatic light in this paper. Our work can be extend to the case of a multi-level atom interacting with multi-frequency light, with which the non-linear effect of the light-atom interacting system can be studied. The quantization of multi-frequency electromagnetic field is
\begin{equation}
A^{\mu}=\sum_k\sum_{\lambda=1}^2\varepsilon\hat{\epsilon}_{k,\lambda}(a^\dag_k e^{-ik\cdot x-\omega_k t}+a_ke^{ik\cdot x+\omega_k t}),\label{QA}
\end{equation}
where $\omega_k=ck$. The quantization of the bound-state Dirac fields in a multi-level atom is
\begin{equation}
\begin{aligned}
\varphi_B&=\sum_p\sum^M_{m=1}\beta_{p,m}\phi_m(x)e^{\frac{i}{\hbar}p\cdot x}e^{\frac{i}{\hbar}E_mt}b_m,\\
\tilde{\varphi}_B&=\sum_p\sum^M_{m=1}\beta^*_{p,m}\phi^*_m(x)e^{-\frac{i}{\hbar}p\cdot x}e^{-\frac{i}{\hbar}E_mt}b^\dag_m,\label{QBD}
\end{aligned}
\end{equation}
where $m$ is the subscript of a energy level, and $M$ is the number of energy levels. An example is the model of a three-level atom ($|g>, |e1>, |e2>$) interacting with a two-frequency ($\omega_1, \omega_2$) light, with $\omega_1$ ($\omega_2$) light causing the transition between $|g>$ and $|e1>$ ($|e2>$). In such case, the time-dependent interaction Hamiltonian in Eq.~(\ref{HI}) can be rewritten as
\begin{equation}
H_I(t)=q_1e^{i\Delta_1t}a_1b^\dag_{e1}b_g+q_2e^{i\Delta_2t}a_2b^\dag_{e2}b_g+H.c.,\label{HII}
\end{equation}
where $\Delta_1=\omega_1-(\omega_{e1}-\omega_g)$, $\Delta_2=\omega_2-(\omega_{e2}-\omega_g)$, and $q_1$ ($q_2$) is the spatial integration in Eq.~(\ref{int}) with $\beta^*_e$ and $\phi^*_e(x)$ being replaced by $\beta^*_{e1}$ ($\beta^*_{e2}$) and $\phi^*_{e1}(x)$ ($\phi^*_{e2}(x)$). After defining the three-level density matrix and the spontaneous emission rate of $|e1>$ and $|e2>$, which are $\Gamma_1$ and $\Gamma_2$ respectively, a set of OBEs can be derived from the interaction Hamiltonian in Eq.~{\ref{HII}} to describe a typical nonlinear spectrum effect, which is the electromagnetically-induced transparency (EIT) \cite{EIT}. Besides, the extensions of our QED-based OBEs with different number of atomic energy levels and light frequency can study other nonlinear spectra of atoms including recoil-induced resonances (RIR) \cite{RIR} and electromagnetically-induced absorption (EIA) \cite{EIA}. Since all these nonlinear spectra are formerly studied by the OBEs with EDA, when EDA becomes invalid the extension of our QED-based OBEs would be a better choice.

The QED-based OBEs also fully describe the interactions between bound electrons and both the electric and magnetic field of light. Besides the dipole approximation, the electric dipole approximation (EDA) also ignores the interaction between the atoms and the magnetic field $\mathbf{B}$ of the light due to that this interaction is much smaller than it between the atoms and the electric field $\mathbf{E}$ of the light. However, the interaction Hamiltonian density in Eq.~(\ref{I-B}) is the \emph{minimal-coupling} in bound-state QED, where the bound-state Dirac field $\varphi_B$ couples the four-potential $A^{\mu}$ of the free electromagnetic field. The four-potential $A^{\mu}=\mathbf{A}+\phi$ contents the contributions from both the electric field of the light by $\mathbf{E}=-\nabla\phi-(1/c)\cdot\partial\mathbf{A}/\partial t$, and the magnetic field of the light by $\mathbf{B}=\nabla\times\mathbf{A}$. Therefore, the bound-state QED based OBEs are beyond the EDA not only by the dipole approximation part, but also by the electric approximation part.

In summary, we derive a new set of optical Bloch equations (OBE) from the minimal-coupling Hamiltonian density between the bound-state Dirac field and the free electromagnetic field in bound-state QED, which is Eq.~(\ref{I-B}). The time-dependent interaction Hamiltonian, Eq.~(\ref{HI}), is the spatial integration of the minimal-coupling Hamiltonian density, which does not have the electric dipole approximation (EDA). Then our new set of OBEs is an useful tool to study the quantum optics of the light-atom interacting system where the wavelength of the light is close to or smaller than the scale of an atom. Since recent realization of X-ray lasers \cite{FEL,X-ray} provide the opportunity to study the quantum optics phenomena of atoms interacting with X-ray experimentally, our new OBEs can predict that a two-level atom with X-ray transition frequency in a X-ray beam will have the absorption spectra which are described by Fig.~\ref{absorption}, and the Rabi oscillation which are described Fig.~\ref{evolution}. The duration of the free-electron laser pump pulse for the atomic X-ray laser can also be optimized by the solutions of the new OBEs. When the new OBEs are extended to the case of a multi-level atom interacting with multi-frequency light, the non-linear effect of such light-atom interacting system can be studied by them. Therefore, we hope our work could be an opening work of developing more OBEs beyond the EDA.

\section*{ACKNOWLEDGEMENT}

This work was supported by the NKBRSFC under grants Nos. 2011CB921502, 2012CB821305, 2009CB930701, 2010CB922904, NSFC under grants Nos. 10934010, 60978019, and NSFC-RGC under grants Nos. 11061160490 and 1386-N-HKU748/10.

\end{document}